\documentclass[letter,twocolumn]{jpsj3}

\usepackage{txfonts}
\usepackage{graphicx}
\usepackage{bm}
\usepackage{color}
\usepackage{amsmath}

\begin{document}

\title{ Theory of Cross-correlated Electron--Magnon Transport Phenomena: \\ 
Case of Magnetic Topological Insulator
}
\date{\today}
\author{Yusuke Imai and Hiroshi Kohno}
\inst{Department of Physics, Nagoya University, Nagoya 464-8602, Japan}
\abst{We study transport phenomena cross-correlated among the heat and electric currents 
of magnons and Dirac electrons on the surface of ferromagnetic topological insulators. 
 For a perpendicular magnetization, we calculate magnon- (electron-) drag anomalous Nernst/Seebeck 
(anomalous Ettingshausen/Peltier) effects and magnon-/electron-drag thermal Hall effects. 
 The magnon-drag 
thermoelectric effects are interpreted to be caused by magnon-induced electromotive force. 
 When the magnetization has in-plane components, 
there arise thermal/thermoelectric analogs of anisotropic magnetoresistance (AMR).  
 In the insulating state, the thermal AMR is realized as a magnonic analog of AMR. 
}

\maketitle

 Three-dimensional topological insulators (TIs) \cite{MorreBalents, FuKane, QiHughesZhang, Hsieh} 
have attracted much attention 
from the viewpoint of spin-related transport. 
 They host characteristic surface states described as two-dimensional (2D) Dirac electrons, 
whose spin direction is locked perpendicular to the wave vector (and surface normal), 
a property called spin-momentum locking \cite{NomuraNagaosa}. 
 Recently, systems in which 2D Dirac electrons coexist with ferromagnetism were realized 
on the surface of magnetic topological insulators (MTIs) 
by doping magnetic impurities in TIs \cite{Yu, ChangXue}.

Various novel transport phenomena are expected on the surfaces of MTIs
because of the interplay between Dirac electrons and magnetization. 
 The experiments reported to date include the quantum anomalous Hall (QAH) effect \cite{ChangXue} 
and unidirectional magnetoresistance \cite{YasudaTokura}. 
 The former is related to the topological index of the (gapped) Dirac band structure, 
and is caused by the {\it perpendicular} component of the magnetization. 
 The latter is caused by the nonreciprocal scattering of Dirac electrons by spin waves (magnons) 
in the presence of an {\it in-plane} component of the magnetization.  
 Such mutual interactions of electrons and magnons are expected to affect their transport properties  
at a more fundamental level.

 In this letter, we report on mutual drag effects of electrons and magnons on the surfaces of MTIs \cite{OkumaNomura}. 
 In drag processes, a current of one kind of particle (electrons or magnons) 
is induced by another kind of particle. 
 They may be viewed as \lq\lq cross-correlation'' effects in the sense that they connect different kinds of currents. 
 We specifically calculate (a) magnon-drag anomalous Nernst/Seebeck effects (MdNE/MdSE), 
 (b) magnon- (electron-) drag electron (magnon) thermal Hall effect (MdTHE, EdTHE), and 
 (c) thermal/thermoelectric analogs of anisotropic magnetroresistance (AMR). 
 A physical picture of (a) will be provided in terms of the magnon-induced electromotive force. 
 While (b) is related to a violation of the Wiedemann--Franz (WF) law, 
(c) is significantly enhanced in the insulating state and realized as a magnonic AMR.


 To study the coupled electron--magnon system on the surface of an MTI, 
we start from the Hamiltonian 
\begin{align}
 H &=  \int d^2x \, c^{\dagger } \biggl\{ - \frac{1}{2} v_{\rm F} 
        \frac{\hbar}{i}({\overleftrightarrow{\bm \nabla}}\times{\bm \sigma})^z 
        + V_{\rm imp}({\bm r}) - J_{sd} \, {\bm S} \cdot {\bm \sigma} \biggr\} \, c 
\nonumber \\
 & + J \int d^2x \, (\partial_i {\bm S})^2, 
\label{eq;H0}
\end{align}
for 2D Dirac electrons, described by the creation/annihilation operators 
 $c^{\dagger}=(c_{\uparrow}^{\dagger},c_{\downarrow}^{\dagger})$ and 
$c=(c_{\uparrow},c_{\downarrow})^{\rm T}$, 
and localized spins, ${\bm S}$, responsible for the ferromagnetic moment. 
 Here $v_{\rm F}$ is the slope of the Dirac cone, 
${\bm \sigma}$ is the Pauli matrix, 
$V_{\rm imp}$ is a random impurity potential, 
$J_{sd}$ is the $s$-$d$ exchange coupling constant, and 
$J$ is the exchange stiffness constant. 
 We introduce magnons according to the Holstein--Primakoff transformation \cite{HolsteinPrimakoff}, 
$S_{x'} - iS_{y'} = r_0 a^{\dagger} (2S - r_0^2 a^{\dagger} a )^{1/2} \simeq \sqrt{2S} r_0 a^{\dagger}$ 
and $S_{z'} = S - r_0^2 a^{\dagger} a$, 
where $z'$ is the equilibrium direction $\langle {\bm S} \rangle_{\rm eq}$ of ${\bm S}$, 
and $r_0$ is the lattice constant. 
 Up to the first order in $S^{-1}$ (or $s_0^{-1}$), the Hamiltonian is rewritten as 
\begin{align}
 &H = H_{\rm el} + H_{{\rm el},{\rm mag}} + H_{\rm mag}, 
\label{eq:H}
\\
 &H_{\rm el} = \sum_{\bm k} c_{\bm k}^{\dagger} 
       [-\hbar v_{\rm F}({\bm k}\times{\bm \sigma})^z - \mu - {\bm M} \cdot {\bm \sigma} ] \, c_{\bm k}, 
\\
 &H_{{\rm el},{\rm mag}} = \int d^2x \left\{ 
   M \sqrt{\frac{2}{s_0}} c^{\dagger}({\Sigma}^+a^{\dagger}+\Sigma^-a) \, c 
   - \frac{M}{s_0} a^{\dagger } a c^{\dagger} \Sigma_z c \right\} , 
\\
 &H_{\rm mag} = \sum_{\bm q} \omega_{\bm q}a_{\bm q}^{\dagger}a_{\bm q},
\end{align}
where $s_0=S/r_0^2$, ${\bm M} = J_{sd} \, \langle {\bm S} \rangle_{\rm eq}$, 
$\Sigma^{\pm} = {\cal R}(\sigma_x \pm i\sigma_y )/2$ 
with an SO(3) matrix ${\cal R}$ that satisfies ${\cal R}{\bm M}=M_z \hat z$, and 
$\omega_{\bm q} = Jq^2+\Delta$ is the magnon dispersion 
with $\Delta$ being a gap. \cite{com_gap} 
 Without loss of generality, we set ${\bm M} = (0, M_y, M_z) = ( 0, M \sin\theta, M \cos\theta)$.


 Assuming point-like impurities, $V_{\rm imp}=u_0\sum_{i}\delta({\bm r}-{\bm r}_i)$, 
and using the Born approximation for the self-energy (Fig.~1 (a)), 
we obtain the Green function of Dirac electrons as \cite{SakaiKohno}
$ G^{\rm R} = (g_0 - {\bm g} \cdot {\bm \sigma})^{-1}$, 
where 
$ g_0 = \epsilon + \mu + i\gamma$, 
$ {\bm g} = (g_x, g_y, g_z) = (\hbar v_{\rm F} k_y , -\hbar v_{\rm F} k_x - M_y ,  -(M_z/\mu) ( \mu-i\gamma))$, 
$\gamma = \gamma_0 |\mu| \, \Theta( |\mu| - |M_z|)$ and
$\gamma_0 = n_iu_0^2 / (2\hbar v_{\rm F})^2$, with the Heaviside step function $\Theta$. 
 For magnons, we consider the self-energy due to the coupling to electrons (Fig.~1 (b)), 
and the Green function is given by 
\begin{align}
 D_{\bm q}^{\rm R}(\nu) = (\nu-\tilde{\omega}_{\bm q} + i\alpha \nu )^{-1} .
\label{eq:D_mag}
\end{align}
 Here 
\begin{align}
 \tilde{\omega}_{\bm q} &= Jq^2 + \Delta + \delta J \, [(1+\sin ^2\theta) \, q_x^2 + q_y^2] , 
\label{eq:tilde_omega_q}
\end{align}
is the renormalized magnon energy\cite{com_DMI}\,; 
the last term introduces the anisotropy in the magnon dispersion 
through the renormalization of the exchange stiffness constant, \cite{WakatsukiNagaosa}  
\begin{align}
  \delta J &= \frac{M^2}{24 \pi s_0 |M_z|} \, \Theta(-|\mu|+|M_z|)  , 
\label{eq;dJ}
\end{align}
which is finite only when the chemical potential $\mu$ lies in the gap, $|\mu | < |M_z|$. 
 This indicates that, in the QAH state, the magnon dispersion is anisotropic 
if the magnetization has an in-plane component. 
 The last term in the denominator of Eq.~(\ref{eq:D_mag}) represents Gilbert damping with 
\begin{align}
  \alpha &= \alpha_d +  \alpha_c \, 
   \left[ 1 - \frac{\mu^2 - 2M_z^2}{2(\mu^2 - M_z^2)} \sin^2\theta \right] 
   \, \Theta( |\mu|-|M_z|)  . 
\label{eq;alpha}
\end{align}
 Here 
$\alpha_c = M^2 (\mu^2-M_z^2) \, 
              / [8 \pi (\hbar v_{\rm F})^2 s_0 (\mu^2+M_z^2) \gamma_0]$
originates from the coupling to Dirac electrons and is present only in the \lq\lq metallic'' state, 
$|\mu|>|M_z|$\cite{SakaiKohno}, 
whereas $\alpha_d$ exists even in the insulating state, $|\mu|<|M_z|$, 
originating from other processes. 
 It is generally expected that $\alpha_d \ll \alpha_c$. 
 We note that the Gilbert form of the damping may not be appropriate for high-energy magnons 
(e.g., at room temperatures). 
 Therefore, in this letter, we restrict ourselves to sufficiently low temperatures (but above the magnon gap).


  To calculate the response of a physical quantity $\hat{A}_i$ to a temperature gradient, 
we follow Luttinger \cite{Luttinger, KohnoBauer} and use the formula, 
$\langle \hat{A}_i \rangle 
 = \lim_{\omega \to 0} (i\omega)^{-1} [K_{ij}(\omega+i0)-K_{ij}(0)] (-\nabla_j T/T) 
 \, \equiv \kappa_{ij} \, (-\nabla_j T)$, 
where 
$K_{ij}(i\omega_{\lambda}) 
 = \int_0^{T^{-1}} d\tau e^{i\omega_{\lambda}\tau}\langle {\rm T}_{\tau}\hat{A}_i(\tau)J^{Q}_j\rangle$ 
with the total energy current $J^Q_i$ of the system. 
 From the continuity equation, the energy current density is obtained as 
$ {\bm j}_{\rm el}^{Q} + {\bm j}_{\rm mag}^{Q}$, where 
\begin{align}
 &{\bm j}_{\rm el}^{Q} = \frac{i\hbar}{2}(c^{\dagger}{\bm v}\dot{c}-\dot{c}^{\dagger}{\bm v}c) , 
 \ \ \  \  {\bm v} = v_{\rm F}\hat{z} \times {\bm \sigma}, 
\label{heat current Dirac electron} 
\\
 &{\bm j}_{\rm mag}^{Q} = -J \{ \dot{a}^{\dagger}({\bm \nabla}a) + ({\bm \nabla}a^{\dagger}) \, \dot{a} \} , 
\label{heat current magnon}
\end{align} 
for Dirac electrons and magnons, respectively.


 Before studying the drag effects, we first consider the transport of magnons and Dirac electrons 
occurring independently. 
 The thermal conductivity of magnons is calculated from 
\begin{align}
 (\kappa_{{\rm mag},{\rm mag}}^{QQ} )_{ij}
= \frac{1}{2\pi T } \sum_{\bm q}u_iu_j \int_{-\infty}^{\infty}d \nu 
   \left(-\frac{\partial n}{\partial\nu}\right)\nu^2D_{\bm q}^{\rm R}(\nu)D_{\bm q}^{\rm A}(\nu), 
\label{magnon conductivity}
\end{align}
where $u_i=2J q_i$ is the velocity of magnons and $n=n(\nu)$ is the Bose distribution function. 
 Up to ${\cal O}(s_0^{-1})$, the longitudinal component, $\kappa_{xx}$, and the in-plane anisotropy, 
$\kappa_{xx-yy} \equiv \kappa_{xx} - \kappa_{yy}$, are calculated as 
\begin{align}
 (\kappa_{{\rm mag},{\rm mag}}^{QQ} )_{xx}
&= \frac{J{\cal S}_{\rm mag} }{\alpha} - \frac{{\cal S}_{\rm mag}}{\alpha_d} \frac{4M^2+3M_y^2}{M^2}  
    \delta J ,
\label{magnon conductivity}
\\
 (\kappa_{{\rm mag},{\rm mag}}^{QQ} )_{xx-yy}
&=  - \frac{{\cal S}_{\rm mag}}{\alpha_d} \frac{2M_y^2}{M^2} \delta J ,
\label{magnon AMR}
\end{align}
where 
${\cal S}_{\rm mag} 
 = - (\partial / \partial T) \, T \sum_{\bm q} \ln (1 - e^{-\omega_{\bm q}/k_{\rm B} T})$ 
is the equilibrium entropy density of magnons \cite{Yamaguchi}. 
 The anisotropy $\kappa_{xx-yy}$ here represents a magnonic analog of AMR.  
 We remark that the renormalization of the exchange-stiffness constant ($\delta J$), 
which is ${\cal O}(s_0^{-1})$, affects 
the longitudinal thermal transport (including thermal AMR) of magnons in the insulating region 
($|\mu| < |M_z|$), 
and such contributions are larger for a smaller gap ($\sim |M_z|$) of Dirac electrons. 
 Diagrammatically, this originates from the self-energy of magnons. 
 Below, we will show that this result is modified by, but survives, the drag contribution 
(i.e., vertex corrections). 
 Finally, we note that, even if we include the self-energy due to the coupling to Dirac electrons,\cite{com_DMI} 
the thermal Hall effect (THE) does not occur as the velocities of magnons are commutative.

 The Hall transport of Dirac electrons is described by \cite{CuclerSinitsyn} 
\begin{align}
 \sigma_{xy} = \frac{e^2}{2h}   \times \left\{
\begin{array}{ll}
 2 M_z|\mu| / (\mu^2+M_z^2)  \ \ \ &  (|\mu|>|M_z|), 
 \\
 {\rm sgn}M_z  & (|\mu|<|M_z|) , 
\label{eq:AHE}
\end{array}
\right.
\end{align}
for the anomalous Hall effect (AHE), and 
\begin{align}
 (\kappa_{\rm el, el}^{QQ} )_{xy} 
&= \frac{1}{e^2T} \int \left(-\frac{\partial f}{\partial\epsilon}\right) \epsilon^2 \sigma_{xy}(\epsilon) 
\simeq  \frac{T}{3}\left(\frac{\pi k_{\rm B}}{e}\right)^2\sigma_{xy} , 
\label{eq:eTHE}
\end{align} 
for anomalous THE. 
 Here, the WF law (for Hall transport) holds when the drag effects 
are neglected \cite{com_WF_law}. 
 The Nernst conductivity is given, for $|\mu|>|M_z|$, by 
\begin{align}
 (\kappa_{\rm el, el}^{EQ} )_{xy} 
&\simeq  - \frac{\pi^2}{3} \frac{T}{e} \sigma_{xy}'(\mu) 
=  \frac{\pi^2}{3} \frac{eT}{h} \frac{(\mu^2 - M_z^2) M_z}{(\mu^2 + M_z^2)^2} {\rm sgn} \mu . 
\label{eq:eNernst}
\end{align}

\begin{flushleft}
\begin{figure}[t]
\begin{center}
  \includegraphics[width=8.5cm]{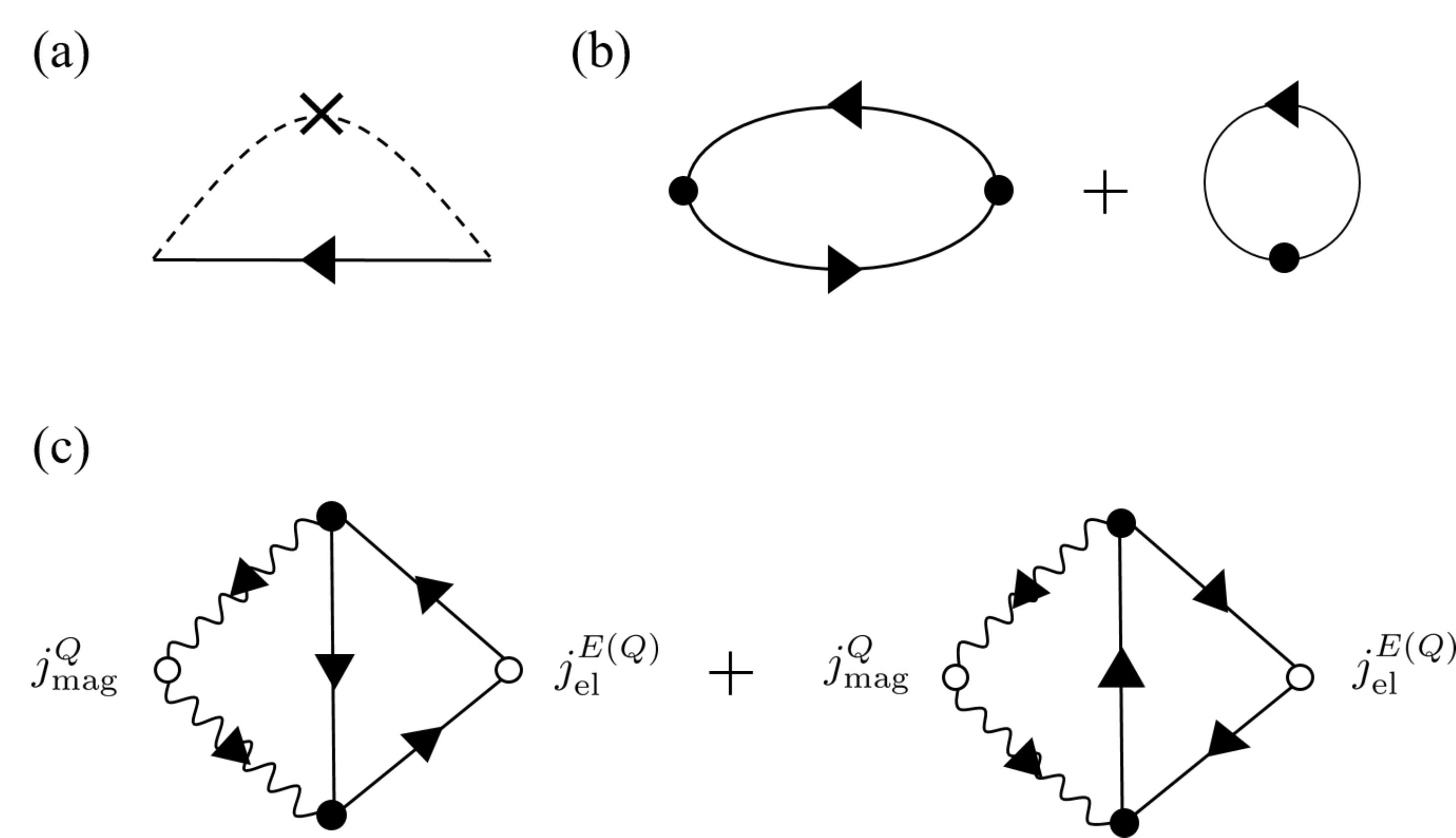}
  \caption{Diagrammatic expression for (a) electron self-energy, (b) magnon self-energy, and 
              (c) Dirac-electron-drag magnon current. 
              The solid (wavy) line with an arrow is the electron (magnon) Green function. 
              The dashed line with a cross is the impurity potential. 
              The filled (empty) circle is the electron--magnon (external) vertex. 
              Diagrams for the magnon-drag electron current are given by (c) with left--right reversed.  } 
  \label{figure drag}
\end{center}
\end{figure}
 \end{flushleft}

 We now study the drag effects.  
 They may be called \lq\lq cross-correlated'' electron--magnon transport 
and are expressed by the correlation functions of two currents, 
one for electrons and the other for magnons, 
$\langle  j_{{\rm mag},i}^Q ; \, j_{{\rm el},j}^{Q \, {\rm or} \, E}  \rangle$ and  
$\langle j_{{\rm el},i}^{Q \, {\rm or} \, E} ; \, j_{{\rm mag},j}^{Q} \rangle$,
where ${\bm j}_{\rm el}^E$ is the electric current. 
 The average $\langle\cdots\rangle$ is defined with the full Hamiltonian $H$ including  
$H_{{\rm el},{\rm mag}}$, which connects the two currents. 
  With respect to the electron--magnon coupling parameter $s_0^{-1}$, the leading contribution is 
${\cal O}(s_0^{-1})$, 
\begin{align}
 \left[ K_{{\rm mag},{\rm el}}^{QQ} (i\omega_{\lambda}) \right]_{ij} 
&= \frac{4\hbar JM^2}{s_0} T\sum_l \sum_{{\bm q}}q_i\left(i\nu_l-\frac{i\omega_{\lambda}}{2}\right) 
\nonumber \\
&\times D_{\bm q}(i\nu_l)D_{\bm q}(i\nu_l-i\omega_{\lambda}) \, 
    {\cal E}_j(i\nu_l;i\omega_{\lambda}), 
\label{Kernel Matsubara} 
\end{align}
given by the diagrams in Fig.~1 (c), 
where $ {\cal E}_j = {\cal E}_j^A + {\cal E}_j^B$,  
\begin{align}
 {\cal E}_j^A &= T\sum_n\sum_{\bm k}\left(i\epsilon_n-\frac{i\omega_{\lambda}}{2}\right) 
\nonumber \\
 &\times {\rm tr} \, [\Sigma^-G_{{\bm k}-{\bm q}}(i\epsilon_n-i\nu_l) 
                             \Sigma^+G_{\bm k}(i\epsilon_n)v_jG_{\bm k}(i\epsilon_n-i\omega_{\lambda})] , 
\\
 {\cal E}_j^B &= T\sum_n\sum_{\bm k}\left(i\epsilon_n+\frac{i\omega_{\lambda}}{2}\right) 
\nonumber \\
 &\times {\rm tr} \, [\Sigma^+G_{{\bm k}+{\bm q}}(i\epsilon_n+i\nu_l) 
                             \Sigma^-G_{\bm k}(i\epsilon_n+i\omega_{\lambda})v_jG_{\bm k}(i\epsilon_n)] .
\end{align}
 The factor $\left(i\epsilon_n\pm\frac{i\omega_{\lambda}}{2}\right)$ originates from the time derivative 
in the heat currents, Eqs.~(\ref{heat current Dirac electron}) and (\ref{heat current magnon}), 
for the response to a temperature gradient ($K^{QQ}$), 
and is absent for the response to an electric field ($K^{QE}$). 
 Similar diagrams have been considered earlier for systems with \cite{OkumaNomura} 
and without \cite{Miura,Yamaguchi} spin-orbit coupling.

 We first consider the case of perpendicular magnetization, ${\bm M} \parallel {\hat z}$. 
 Owing to the perpendicular component $M_z$, the Dirac electrons exhibit AHE (see Eq.~(\ref{eq:AHE})). 
 Therefore, Hall-like (transverse) transport may also be expected in drag processes. 
 The results of ${\cal O}(s_0^{-1})$ are summarized as follows:
\begin{align}
 (\kappa_{\rm el,mag}^{EQ})_{xy-yx} 
&=  \frac{{\cal S}_{\rm mag} }{8\pi\alpha} 
     \frac{M_z^2}{s_0}  
     \frac{eM_z\mu}{(\mu^2+M_z^2)^2 \gamma_0} = (\kappa_{\rm mag,el}^{QE})_{xy-yx} , 
\label{MdNE} 
\\
 (\kappa_{\rm el, mag}^{QQ}) _{xy-yx}
&= - \frac{\pi  T^2 {\cal S}_{\rm mag}}{24\alpha} 
       \frac{M_z^2}{s_0}  
       \frac{(3\mu^2-M_z^2)M_z}{(\mu^2+M_z^2)^3 \gamma_0} = (\kappa_{\rm mag, el}^{QQ})_{xy-yx},
\label{dTHE} 
\\
 (\kappa_{\rm el,mag}^{EQ})_{xx}
&= \frac{ {\cal S}_{\rm mag} }{32 \pi\alpha}\frac{M^2}{s_0} 
      \frac{ e \mu^2 (\mu^2-M_z^2) }{(\mu^2+M_z^2)^3\gamma_0^2} 
= (\kappa_{\rm mag,el}^{QE})_{xx},
\label{MdSE}
\\
 (\kappa_{\rm el, mag}^{QQ})_{xx}^{\delta J} 
&= \frac{2 {\cal S}_{\rm mag} }{\alpha_d} \, \delta J
= (\kappa_{\rm mag, el}^{QQ})_{xx}^{\delta J}, 
\label{drag J}
\end{align}
where the anti-symmetric part, 
$\kappa_{xy-yx} \equiv (\kappa_{xy} - \kappa_{yx})/2$, represents the Hall component. 
 The conductivity $(\kappa_{\rm el,mag}^{EQ})_{xy-yx}$ represents the 
MdNE, and 
 $(\kappa_{\rm el,mag}^{EQ})_{xx}$ represents the MdSE. 
 The contribution to thermal conductivity, 
$(\kappa_{\rm el, mag}^{QQ} + \kappa_{\rm mag, el}^{QQ})_{xx}^{\delta J}$, 
proportional to $\delta J$ is to be added to Eq.~(\ref{magnon conductivity}).

 As indicated in Eq.~(\ref{MdNE}), the MdNE and the 
electron-drag Ettingshausen effect are Onsager partners 
having the same conductivity (with the same sign). 
 The same holds for the MdTHE and the EdTHE (see Eq.~(\ref{dTHE})), 
which constitute the drag contribution to THE and 
violate the WF law for Hall transport, cf. Eq.~(\ref{eq:eTHE}).

 The results (\ref{MdNE})--(\ref{drag J}) are expressed as products of three factors. 
 The first factor (${\cal S}_{\rm mag}/\alpha$) originates from magnons, 
the second ($M^2/s_0$) from the electron--magnon coupling, 
and the third from Dirac electrons. 
 (In Eq.~(\ref{drag J}), use Eq.~(\ref{eq;dJ}) for $\delta J$.)

 The magnon factor, $ {\cal S}_{\rm mag}/\alpha$, which is common to Eqs.~(\ref{MdNE})-(\ref{drag J}), 
 is related to the longitudinal component 
of thermal conductivity, Eq.~(\ref{magnon conductivity}). 
 This indicates that magnons propagate straight (i.e., do not bend) even in the anti-symmetric 
($\kappa_{xy-yx}$) drag processes. 
 The dependence on $T$ through ${\cal S}_{\rm mag}$ (and on $\mu$ and $M_z$), 
together with the deviation from the WF law, 
can be key features to discriminate the drag effects experimentally.

 As an illustration, let us consider the Nernst effect. 
 If the magnons are 2D, ${\cal S}_{\rm mag} \propto T$ at low $T$ (but above $\Delta$), 
in contrast to 3D\cite{Watzman}. 
 Thus, the drag Nernst conductivity, 
$(\kappa_{\rm el, mag}^{EQ})_{xy-yx} \propto {\cal S}_{\rm mag}$ [Eq.~(\ref{MdNE})], 
has the same $T$-dependence as that from a purely electronic process, 
$(\kappa_{\rm el, el}^{EQ})_{xy}$ [Eq.~(\ref{eq:eNernst})], but the magnitude is much larger. 
 In fact, with \cite{SakaiKohno}  
$|M_z| = 0.1 \sim 0.01$eV, $r_0=0.5$nm, and $|M_z|/J_{\rm ex} \sim 1$, where $J_{\rm ex} \equiv J/r_0^2$, 
the ratio 
\begin{align*}
 R \equiv \frac{(\kappa_{\rm el, mag}^{EQ})_{xy-yx}}{(\kappa_{\rm el,el}^{EQ})_{xy-yx}} 
= \frac{|M_z|}{16\pi J_{\rm ex}} \left( \frac{\lambda_M}{r_0} \right)^2 
   \frac{|\mu M_z| (\mu^2 + M_z^2)}{(\mu^2 - M_z^2)^2} 
\end{align*} 
($\lambda_M \equiv 2\pi \hbar v_{\rm F}/|M_z|$)
assumes $R = 10 \sim 1000$ for $\mu = 2M_z$, and $R = 1 \sim 100$ for $\mu = 10 M_z$. 
 This indicates that the Nernst effect is dominated by the drag process. 
 Moreover, one can show that it is related to the deviation from the WF law, 
$\Delta \kappa \equiv  (\kappa_{\rm el, mag}^{QQ}) _{xy-yx} + (\kappa_{\rm mag, el}^{QQ}) _{xy-yx}$, as 
\begin{align}
 (\kappa_{\rm el,mag}^{EQ})_{xy-yx} 
&=  - \frac{3}{2\pi^2}  
       \frac{ e \mu (\mu^2+M_z^2)}{(3\mu^2-M_z^2)} \, \frac{\Delta \kappa }{T^2} . 
\label{eq:relation} 
\end{align} 
 This relation may be tested by experiments.

\begin{figure}[t]
\begin{center}
  \includegraphics[width=6.0cm]{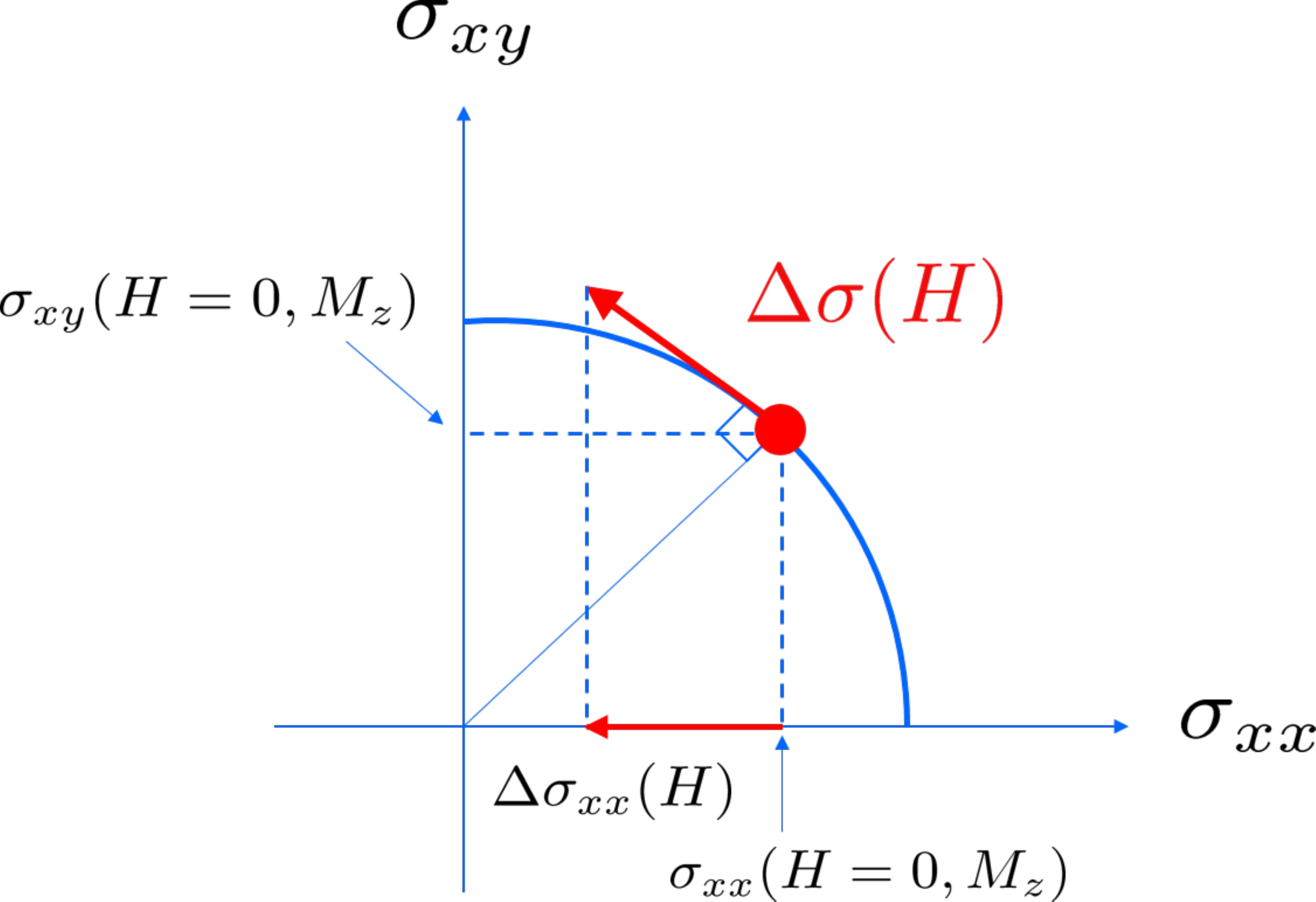}
  \caption{(Color online) Longitudinal magnetoconductance $\Delta \sigma_{xx}(H)$ 
  expressed in the conductivity plane $(\sigma_{xx}, \sigma_{xy})$. 
  The dot on a circle represents the anomalous Hall state with finite $M_z$ in zero magnetic field, $H=0$. 
  The arrow indicated by $\Delta \sigma (H)$ represents the contribution of normal Hall conductivity 
  induced by an applied field $H$.}
  \label{figure longitudinal magnetoconductance}
\end{center}
\end{figure}

 The electron part of Eqs.~(\ref{MdNE}) and (\ref{MdSE}) can be understood as follows.
\cite{com_Mott} 
 In Eq.~(\ref{MdNE}), the electron contribution is $eM_z^3\mu/[\gamma_0(\mu^2+M_z^2)^2]$. 
 This factor also appears in the longitudinal magnetoconductance \cite{SakaiKohno}, 
\begin{align}
 \langle \, {\bm j}_{\rm el}^E \, \rangle 
&= \mp ep^2 (A - B \, \hat z \, \times \, ) ( {\bm E} + {\bm E}^{\rm eff} ) (H + H^{\rm eff})  
\label{MR}
\end{align} 
through the coefficient, $A \equiv - M_z^3\mu / [ 4\pi\gamma_0 (\mu^2+M_z^2)]$, 
where $p \equiv ev_{\rm F}/M$. 
 This relation has been derived by assuming a (classical) small-amplitude deviation ${\bm u}$ 
from the uniformly magnetized state ${\bm n} = \pm \hat z$ 
such that ${\bm n} = \pm \hat{z} + {\bm u}$ with $|{\bm u}| \ll 1$. 
 Here
${\bm A}^{\rm eff} = -p^{-1}{\hat z}\times{\bm u}$, 
${\bm E}^{\rm eff} = -\partial {\bm A}^{\rm eff} / \partial t$, and
$H^{\rm eff} = ({\rm rot}{\bm A}^{\rm eff})_z$ 
are the effective electromagnetic fields (vector potential, electric field, and magnetic field, respectively) 
experienced by the Dirac electrons. 
 The real electromagnetic fields are denoted by ${\bm E}$ and $H$. 
 The term $\sim A {\bm E} H$ in Eq.~(\ref{MR}) corresponds to the (real) magnetoconductance 
induced by an applied magnetic field ${\bm H} = (0,0,H)$, 
which exists at the linear order in $H$ because of the presence of AHE already at $H=0$ 
(see Fig.~\ref{figure longitudinal magnetoconductance}). 
 In fact, the changes in the conductivity tensor caused by $H$,  
$\Delta \sigma_{xx} (H)$ $(\propto A)$ and $\Delta \sigma_{xy} (H)$ $(\propto B)$, 
satisfy the relation \cite{SakaiKohno}, 
\begin{align}
  \Delta \sigma_{xx} (H) &= - \alpha_{\rm AHE} \, \Delta \sigma_{xy} (H)  
\simeq - \alpha_{\rm AHE} \, \sigma_{xy} (H,0)  , 
\end{align}
where $\alpha_{\rm AHE} = \sigma_{xy}(0,M_z)/\sigma_{xx}(0,M_z)$ 
is the anomalous Hall angle and  
$\sigma_{ij}(H,M_z)$ is the conductivity tensor in the presence of $M_z$ and $H$. 
 This shows that the transverse current in the MdNE 
(determined by $A \propto \Delta \sigma_{xx} (H)$) is caused by the AHE of Dirac electrons, 
i.e., it originates at the electron side.

 The relation (\ref{MR}) offers a more direct explanation of Eq.~(\ref{MdNE}) based 
on the effective electromotive forces generated by magnetization dynamics.\cite{phenomenology}  
 To see this, we focus on the term 
$\sim {\bm E}^{\rm eff}H^{\rm eff} \propto ({\hat z} \times \dot {\bm u}) \, {\rm div} {\bm u}$ 
in Eq.~(\ref{MR}) 
and rewrite it using magnon operators $u_x+iu_y \to (2/s_0)^{1/2} a$. 
 Interestingly, it can be expressed by the magnon heat current ${\bm j}_{\rm mag}^{Q}$ 
[Eq.~(\ref{heat current magnon})] as 
$\langle j_{{\rm el},x}^E \rangle 
 = (e/2s_0J)[A\langle j_{{\rm mag},y}^{Q} \rangle - B\langle j_{{\rm mag},x}^Q \rangle]$ 
apart from total derivative terms, $ \sim d (\cdots )/dt$ and $\nabla (\cdots )$. 
 Therefore, a magnon heat current 
$\langle {\bm j}_{\rm mag}^{Q} \rangle = (\kappa_{\rm mag, mag}^{QQ})_{xx}(-\nabla T)
 = (J {\cal S}_{\rm mag}/ \alpha)(-\nabla T)$ 
induced by a temperature gradient also indicates the induction of an electric current, 
\begin{align}
 \langle j_{{\rm el},x}^E \rangle = \frac{ e {\cal S}_{\rm mag} }{2\alpha s_0} 
   \left\{ A \, ( - \nabla_y T) - B \, ( - \nabla_x T ) \right\}  .
\label{drag pheno}
\end{align} 
 This exactly reproduces the microscopic result, Eq.~(\ref{MdNE}), for the MdNE. 
 This also reproduces the MdSE, Eq.~(\ref{MdSE}), but there appears to be a slight disagreement 
by a factor of $\mu^2/(\mu^2+M_z^2)$.

 When the magnetization ${\bm M}$ is tilted from the $\pm \hat z$ direction 
and has an in-plane component, $M_y$, 
the in-plane conduction becomes anisotropic and various kinds of AMR arise. 
 They are calculated as follows, 
\begin{align}
  (\kappa_{{\rm mag},{\rm el}}^{QE})_{xy+yx}
&= \frac{3 {\cal S}_{\rm mag} }{128\pi\alpha} \frac{M_y^2}{s_0} 
    \frac{eM_z\mu}{(\mu^2+M_z^2)^2\gamma_0} 
= -(\kappa_{\rm el,mag}^{EQ})_{xy+yx} ,
\label{drag planar Hall E}
\\
  (\kappa_{{\rm mag},{\rm el}}^{QQ})_{xy+yx}
&= \frac{\pi  T^2 {\cal S}_{\rm mag} }{128\alpha} 
     \frac{M_y^2}{s_0} \frac{M_z(M_z^2-3\mu^2)}{(\mu^2+M_z^2)^3\gamma_0}
= -(\kappa_{\rm el,mag}^{QQ})_{xy+yx} ,
\label{drag planar Hall Q} 
\\
(\kappa_{{\rm mag},{\rm el}}^{QE})_{xx-yy} 
&= \frac{3 {\cal S}_{\rm mag} }{32\pi\alpha}\frac{M_y^2}{s_0} 
     \frac{e(\mu^2+M_z^2)\gamma_0}{(\mu^2-M_z^2)^2} 
= (\kappa_{\rm el,mag}^{QE})_{xx-yy},
\label{drag AMR E} 
\\
(\kappa_{{\rm mag},{\rm el}}^{QQ})_{xx-yy}
&= \frac{\pi T^2 {\cal S}_{\rm mag} }{16\alpha} \frac{M_y^2}{s_0} 
     \frac{M^2\mu(\mu^2+3M_z^2)\gamma_0}{(\mu^2-M_z^2)^3} 
=(\kappa_{\rm el,mag}^{QQ})_{xx-yy}, 
\label{drag AMR Q} 
\\
 (\kappa_{\rm mag,el}^{QQ})_{xx-yy}^{\delta J}
&= \frac{ {\cal S}_{\rm mag} }{\alpha_d } \frac{2M_y^2}{M^2} \, \delta J 
= (\kappa_{\rm el,mag}^{QQ})_{xx-yy}^{\delta J} , 
\label{drag AMR J}
\end{align}
where $\kappa_{xy+yx} \equiv (\kappa_{xy} + \kappa_{yx})/2$ 
and $\kappa_{xx-yy} \equiv \kappa_{xx} - \kappa_{yy}$ reflect the in-plane anisotropy. 
 These effects may be called cross-correlated AMR or drag AMR (both thermal and thermoelectrical), 
which are caused by the electron-/magnon-drag processes and disappear 
when $M_y=0$.\cite{com_gauge,ChibaBauer}

 All the results obtained above are consistent with the Onsager's theorem,  
$ (\kappa_{\rm mag, el}^{Q, E/Q})_{ij\pm ji}({\bm M}) 
 = \pm (\kappa_{\rm el, mag}^{E/Q, Q})_{ij\pm ji}(-{\bm M})$,  
where the upper sign is for the symmetric part (longitudinal and planar Hall conductivity), 
and the lower sign is for the anti-symmetric part (Hall conductivity). 
 As a consequence, 
we have a finite drag contribution, 
$\kappa_{\rm el, mag}^{QQ} + \kappa_{\rm mag, el}^{QQ}$, 
for the symmetric (anti-symmetric) part if it is even (odd) under ${\bm M} \to - {\bm M}$. 
 This is the case for the drag THE [Eq.~(\ref{dTHE})] 
and the drag longitudinal conductivity [Eq.~(\ref{drag J})] 
including the drag AMR [Eqs.~(\ref{drag AMR Q}) and (\ref{drag AMR J})]. 
 In contrast, for the drag planar thermal Hall conductivity [Eq.~(\ref{drag planar Hall Q})], 
the Onsager partners have opposite signs and cancel each other, 
$(\kappa_{{\rm mag},{\rm el}}^{QQ} + \kappa_{{\rm el},{\rm mag}}^{QQ})_{xy+yx} = 0$; 
they do not contribute to the total conductivity.

 In the undoped case, $|\mu | < |M_z|$, the magnon propagation becomes anisotropic because of 
the $\delta J$-term [Eq.~(\ref{eq:tilde_omega_q})]. 
 The total anisotropy of the thermal conductivity is obtained by adding the drag contribution,  
Eq.~(\ref{drag AMR J}), to Eq.~(\ref{magnon AMR}), which gives  
$ \kappa^{QQ}_{xx-yy}  = ( {\cal S}_{\rm mag} /\alpha_d)(2 M_y^2/M^2) \, \delta J$. 
 The \lq\lq AMR ratio'' is 
\begin{align}
 \left( \frac{\kappa^{QQ}_{xx} - \kappa^{QQ}_{yy}}{\kappa^{QQ}_{xx}} \right)_{|\mu | < |M_z|}
&= \frac{M_y^2}{12 \pi s_0 J |M_z|} 
	= \frac{J_{sd}}{12 \pi  J_{\rm ex} } \frac{\sin^2 \theta}{ |\cos\theta \, | }    , 
\label{drag AMR J_total_ratio}
\end{align}
where $J_{\rm ex} = J/r_0^2$ is the ferromagnetic exchange constant (with the dimension of energy), 
and Eq.~(\ref{eq;dJ}) has been used. 
 As $J_{sd}$ is comparable to or larger than $J_{\rm ex}$, the anisotropy ratio 
(\ref{drag AMR J_total_ratio}) is well in a range of experimental detection.

 To summarize, we analytically studied the cross-correlated electron--magnon transport 
phenomena on the surface of MTIs. 
 We first considered the case of perpendicular magnetization, ${\bm M}\parallel{\hat z}$, 
and calculated various drag conductivities of ${\cal O}(s_0^{-1})$, which include  
(i) electron- (magnon-) drag Ettingshausen (Nernst) effect, 
(ii) electron- (magnon-) drag magnon (electron) thermal Hall effect, and 
(iii) electron- (magnon-) drag Peltier (Seebeck) effect. 
 Our microscopic results for the magnon-drag Nernst effect, (i), and 
magnon-drag Seebeck effect, (iii), could be physically interpreted to be caused by
the electromotive force induced by magnon-current magnetization dynamics. 
 The presence of (ii) indicates the violation of the WF law owing to the drag effects.

 For the \lq\lq tilted'' case, ${\bm M}=M_y{\hat y}+M_z{\hat z}$, 
we found (iv) thermoelectric/thermal AMR caused by drag processes. 
 For this AMR, the magnons play an essential role for the Dirac electrons to see    
the in-plane magnetization component. 
 All the effects studied here apply to the \lq\lq doped'' case $|\mu| > |M_z|$ 
except for (iv), which is present also in the insulating case $|\mu| < |M_z|$ 
and gives rise to magnonic AMR. 
 In addition, we observed that the electron-drag planar thermal Hall effect (pTHE) and 
the magnon-drag pTHE cancel each other out in the total THE, 
as assured by Onsager's theorem.

\begin{acknowledgment}
 We would like to thank T.~Yamaguchi for valuable and stimulating discussion. 
 We also thank A. Yamakage, K. Nakazawa, T. Funato, and J. Nakane for daily discussions.  
 This work was supported by JSPS KAKENHI Grant Numbers 25400339, 15H05702, and 17H02929. 
\end{acknowledgment}


\begin{thebibliography}{99}
\bibitem{com_gap} 
 The gap may originate from magnetic anisotropy or applied magnetic field, which are 
 not explicitly written in Eq.~(\ref{eq;H0}). 
 It may also originate from the self-energy,\cite{NunezRossier} 
 but here we denote it by $\Delta$ without invoking its origin.
\bibitem{com_DMI} 
 Precisely speaking, the magnon self-energy allows a term linear in ${\bm q}$ 
corresponding to the Dzyaloshinskii--Moriya interaction.\cite{WakatsukiNagaosa}  
  However, it does not lead to the Hall effect and is thus neglected in this letter.
\bibitem{com_WF_law} 
 Note that the WF law does not hold for the longitudinal conductivity  
 even without the drag effects because of the contribution from magnons. 
 On the other hand, magnons do not exhibit the Hall effect at this level (without drag).
\bibitem{com_Mott} 
 As for Eq.~(\ref{dTHE}), it is related to Eq.~(\ref{MdNE}) via the Mott relation, 
 namely, it can be obtained by applying $[(\pi k_{\rm B}T)^2/ 3 e] (\partial/\partial \mu )$ 
 to Eq.~(\ref{MdNE}).
\bibitem{com_gauge} 
 It may appear evident that the presence of $M_y$ breaks the in-plane 
rotational symmetry and causes AMR. 
 Note, however, that if $M_y$ is uniform and static, there is no AMR, 
because $M_y$ is thereafter a pure gauge degree of freedom invisible to Dirac electrons.
 Magnons introduce spatial/temporal variations and make $M_y$ visible.  
 Mathematically, the expression of the electron--magnon coupling contains the information of $M_y$.
\end{thebibliography}
\end{document}